\documentclass[reprint, superscriptaddress, nofootinbib, amsmath,amssymb, aps, prd]{revtex4-2}

\usepackage{graphicx}
\usepackage{dcolumn}
\usepackage{bm}
\usepackage{amssymb}
\usepackage{diagbox}
\usepackage[x11names]{xcolor}
\usepackage[colorlinks=true,linkcolor=blue,citecolor=Blue4]{hyperref}

\usepackage[T1]{fontenc}

\usepackage{tensor}

\def \H {\mathcal{H}}
\def \B {\mathcal{B}}
\newcommand{\dd}{{\rm{d}}}
\renewcommand{\d}{\mathrm{d}} 

\newcommand{\boldu}{\mbox{\boldmath$u$}} 
\newcommand{\bolde}{\mbox{\boldmath$e$}} 
\newcommand{\boldk}{\mbox{\boldmath$k$}} 
\newcommand{\boldl}{\mbox{\boldmath$l$}} 
\newcommand{\boldm}{\mbox{\boldmath$m$}} 
\newcommand{\boldZ}{\mbox{\boldmath$Z$}} 

\newcommand{\rh}{r_\mathrm{h}}
\newcommand{\sgnL}{\mathrm{sgn} \Lambda \,}
\newcommand{\X}{{\rho}}

\newcommand{\sk}[1]{{\color{blue} #1}}

\newcommand{\vect}[1]{\bm{#1}}

\begin{document}

\preprint{APS/123-QED}

\title{Exact Bachian singularity in quadratic gravity}

\author{{\v S}imon Kno{\v s}ka}
\email{simon.knoska@matfyz.cuni.cz}
\affiliation{
Charles University, Faculty of Mathematics and Physics, Institute of Theoretical Physics,\\
V Hole{\v s}ovi{\v c}k{\' a}ch 2, 180 00 Prague 8, Czech Republic
}
\affiliation{Astronomical Institute of the Czech Academy of Sciences,
Bo\v{c}n\'i II 1401/1a, 141 00 Prague 4, Czech Republic}

\author{David Kofro{\v n}}%
\email{david.kofron@matfyz.cuni.cz}
\affiliation{
Charles University, Faculty of Mathematics and Physics, Institute of Theoretical Physics,\\
V Hole{\v s}ovi{\v c}k{\' a}ch 2, 180 00 Prague 8, Czech Republic
}
\affiliation{Astronomical Institute of the Czech Academy of Sciences,
Bo\v{c}n\'i II 1401/1a, 141 00 Prague 4, Czech Republic}

\author{Robert {\v S}varc}
\email{robert.svarc@matfyz.cuni.cz}
\affiliation{
Charles University, Faculty of Mathematics and Physics, Institute of Theoretical Physics,\\
V Hole{\v s}ovi{\v c}k{\' a}ch 2, 180 00 Prague 8, Czech Republic
}

\date{\today}

\begin{abstract}
For specifically coupled values of the quadratic gravity parameters, we present a fully explicit static spherically symmetric solution. It contains the central singularity surrounded by the black-hole or the cosmological horizon for the negative or positive cosmological parameter, respectively. This spacetime, thus, belongs to the already analyzed classes of solutions expressed in terms of the Frobenius expansions, continuous fractions, or numerical simulations; however, it has not been explicitly identified before. The purpose of the presented highly-constrained somehow unphysical model is to reveal a global geometric picture that may occur in spherical spacetimes within quadratic gravity, which is typically hidden in more general approximative solutions. At the same time, it can serve as a solid benchmark for evaluating the accuracy of the methods employed to obtain such solutions.
\end{abstract}

\maketitle

\section{Introduction}
Modified theories of gravity have garnered considerable attention over the past decade \cite{Sotiriou:2010, DeFelice:2010, Capozziello:2011, Clifton:2012}. This follows from both practical reasons for explaining observational phenomena \cite{Will:2014} and theoretical reasons for discovering mysteries of the quantum theory of gravity. For the latter case, it is natural to allow the quadratic terms in curvature to be present in the action because these terms naturally arise from the semi-classical limit of quantum string-like theories. Additionally, such action is renormalizable with the drawback of introducing ghosts into the theory \cite{Stelle:1977}.

The physically reasonable and natural setting to test the consequences of quadratic gravity is to assume a static and spherically symmetric spacetime. This allows us to study generalizations of the classic Schwarzschild geometry~\cite{Schwarzschild:1916}. Surprisingly, in contrast to the Birkhoff theorem in general relativity (GR), many other \mbox{(non-)black-hole} solutions are present in this setup \cite{LuPerkinsPopeStelle:2015, LuPerkinsPopeStelle:2015b, HoldomRen:2017, Kokkotas:2017, PSPP18, SvarcPodolskyPravdaPravdova:2018, PSPP20, PravdaPravdovaPodolskySvarc:2021} in addition to the Schwarzschild one. Moreover, even with the restrictive spherical assumption, the generalized non-Schwarzschild solutions with the quadratic contribution present cannot be typically expressed in terms of elementary functions, and various techniques to obtain such solutions are employed, namely, the Frobenius-like analysis of the power series, numerical integration, or linearization of the equations in the weak-field regime. 

 One of the most important questions discussed is the asymptotic behavior of the black-hole solutions and whether the asymptotically flat (black-hole) solution with the Schwarzschild limit can even exist. This was investigated, e.g., in \cite{Nelson:2010, LuPerkinsPopeStelle:2015, LuPerkinsPopeStelle:2015b, Daasetal:2023}, and others. It was shown that asymptotically flat black holes are very special. In fact, they require some fine-tuning of the parameters.
 
 As we have already mentioned, quadratic theory in the spherically symmetric regime is richer than in GR and allows for qualitatively different types of solutions. This means that another important task is to identify and classify all possible solutions. The detailed classification was carried out using the conformal-to-Kundt coordinates \cite{PravdaPravdovaPodolskySvarc:2017} in \cite{PravdaPravdovaPodolskySvarc:2021}.  These coordinates offer a surprisingly simple formulation of the quadratic gravity field equations. However, the classification is not necessarily complete because it is based on power series expansions, which are coordinate-dependent. In particular, some coordinates do not allow the power series expansion of a specific solution. Moreover, the Frobenius-like solutions are limited by the convergence radii of the series and can thus describe only the local behavior of the geometry rather than its global structure.
 
Due to the complexity of quadratic gravity and the richness of its solution space, even in the simple case of spherical symmetry \cite{SilveravalleZuccotti:2023}, its deeper understanding remains an open problem. Therefore, the analysis of a fully explicit global exact solution, which will be carried out in this paper, seems very interesting, even though the price to pay is a very specific coupling of the theory constants.

The paper is organized as follows: in the remaining part of the introduction, we summarize the description of quadratic gravity and spherically symmetric geometries. In Section~\ref{Sec:solution} we present the metric form of the Bachian singularity together with its geometrical and physical properties, and in Section~\ref{Sec:Connection} we relate this solution to previous works.

\subsection{Quadratic gravity}

The action of generic quadratic gravity theory in four dimensions \cite{Salvio:2018} should contain possible quadratic curvature terms $R^2, R_{ab}R^{ab}$ and $R_{abcd}R^{abcd}$, that is, the Ricci scalar squared, the Ricci tensor squared, and the Riemann tensor squared, respectively. However, not all these scalars are independent since in $D=4$ the Gauss--Bonnet term is topological invariant. Furthermore, the square of the Weyl tensor can be introduced instead. This means that generic four-dimensional quadratic gravity can be described with the action \cite{Salvio:2018} of the form\footnote{The combination ${- \alpha C_{abcd}C^{abcd} + \beta R^2}$ can be replaced by ${+\tilde{\alpha}R_{ab}R^{ab} + \tilde{\beta}R^2}$ with $\alpha=-\frac{1}{2}\tilde{\alpha}$ and $\beta=\tilde{\beta}+\frac{1}{3}\tilde{\alpha}$, where the triviality of the Gauss--Bonnet term is used.}
\begin{equation}
    S = \int \d^4 x \, \sqrt{-g} \left[ \gamma(R- 2\Lambda) - \alpha C_{abcd}C^{abcd} + \beta R^2 \right] \,, \label{eq:quadratic_gravity_action}
\end{equation}
 with $\gamma^{-1} = G$. It was shown \cite{Stelle:1978} that the Weyl squared term corresponds to the massive spin-2 propagation and the Ricci scalar squared corresponds to the massive scalar propagation. The variation ${\delta S=0}$ of the action results in the field equations
\begin{align}
	&\gamma\left(R_{ab}-\frac{1}{2}Rg_{ab} + \Lambda g_{ab}\right)  \label{eq:4D_EOM}  \\
	&\hspace{3mm}- 4 {\alpha} B_{ab} + 2 {\beta} \left(R_{ab} -\frac{1}{4}Rg_{ab} + g_{ab} \Box - \nabla_a \nabla_b \right)R = 0 \, , \nonumber
\end{align}
with the Bach tensor $B_{ab}$ defined as
\begin{align}
	B_{ab} \equiv \left(\nabla^c \nabla^d + \frac{1}{2}R^{cd}\right)C_{acbd} \, .
	\label{eq:Bach_tensor_Def}
\end{align}
The Bach tensor is traceless (${g^{ab}B_{ab}=0}$), symmetric (${B_{ab}=B_{ba}}$), conserved (${\nabla^b B_{ab}=0}$), and scales under conformal transformation ${g_{ab}=\Omega^2 \tilde g_{ab}}$ as
\begin{equation}
B_{ab}=\Omega^{-2}\tilde B_{ab} \,. \label{BachConformalScaling}
\end{equation}

Furthermore, the field equations can be simplified by assuming the natural condition $R = \mathrm{const}$. This setting is often discussed because, in the specific case of vanishing cosmological constant, $\Lambda = 0$, asymptotically flat solutions containing a horizon must satisfy the $R = 0$ condition throughout the spacetime \cite{LuPerkinsPopeStelle:2015b}. The non-vanishing cosmological constant within the class $R = \mathrm{const}$ implies
\begin{equation}
    R = 4 \Lambda \,, \label{eq:trace_general}
\end{equation}
and the field equations then become
\begin{equation}
    (\gamma +8 \Lambda \beta)\left(R_{ab}- \Lambda g_{ab}\right) = 4 \alpha B_{ab}\,. \label{eq:4D_EOM_constrained}
\end{equation}
There are two branches of solutions depending on the parameters of theory. Either the parameters are fine-tuned, that is ${\gamma +8 \Lambda \beta = 0}$ and ${B_{ab} = 0}$ are simultaneously satisfied \cite{PravdaPravdovaPodolskySvarc:2017}, or there is a generic case $\gamma +8 \Lambda \beta \neq 0$ with
\begin{equation}
    R_{ab} - \Lambda g_{ab} = 4kB_{ab} \label{eq:field_eq}\,,
\end{equation}
where
\begin{equation}
k = \frac{\alpha}{\gamma + 8\beta\Lambda}\neq0 \,. \label{k_definition}
\end{equation}
We focus on this second branch of quadratic theories with the field equations (\ref{eq:field_eq}). 

\subsection{Spherically symmetric geometries}

The classic form of a general static spherically symmetric geometry is
\begin{equation}
\dd s^2 = -h(\bar r)\,\dd t^2+\frac{\dd \bar r^2}{f(\bar r)}+\bar r^2(\dd \theta^2+\sin^2\theta\,\dd \phi^2) \,, \label{metric:SchwCoord}
\end{equation}
where $h(\bar r)$ and $f(\bar r)$ are arbitrary functions of a surface radial coordinate $\bar r$. Vacuum Einstein's general relativity with a cosmological constant then implies
\begin{equation}
h(\bar{r}) = f(\bar{r})=1-\frac{2m}{\bar{r}} -\frac{\Lambda}{3}{\bar r}^2 \,, \label{sol:SchwAdS_SchwCoord}
\end{equation}
which corresponds to the famous Schwarzschild--(anti-)de Sitter spacetime.

However, in the context of quadratic gravity with constant scalar curvature, where the most complicated  ingredient of the field equations (\ref{eq:field_eq}) is the Bach tensor, it is convenient to express the spherical line element in an explicitly conformal form with a simpler seed metric \cite{PravdaPravdovaPodolskySvarc:2021, PravdaPravdovaPodolskySvarc:2017}. This can be achieved using
\begin{equation}
\bar{r} = \Omega(r)\,, \qquad t = u - \int\! \frac{\dd r}{\H(r)} \,, \label{trans:SchwToConfKundt}
\end{equation}
and the identification
\begin{equation}
h\equiv -\Omega^2\, \H \,, \qquad f \equiv -\left(\frac{\Omega'}{\Omega}\right)^2 \H \,, \label{rcehf}
\end{equation}
with the prime denoting the differentiation with respect to $r$. This leads to the conformal-to-Kundt form of the spherical line element
\begin{equation}
\hspace{-3.0mm} \dd s^2 = \Omega^2(r)\! \left[\dd \theta^2+\sin^2\theta\,\dd \phi^2 -2\,\dd u\,\dd r+{\cal H}(r)\,\dd u^2 \right]\!. \label{metric:ConfKundt}
\end{equation}
This parameterization admits the following gauge freedom
\begin{equation}
 r \rightarrow  \lambda^{-1} r + \nu \,, \quad  u \rightarrow \lambda u \,,
\quad \mbox{with} \quad {\cal H} \rightarrow \lambda^{2}{\cal H} \,. \label{GaugeTrans}
\end{equation}
Here, the Schwarzschild--(anti-)de Sitter solution can be expressed as
\begin{equation}
\Omega(r)=-\frac{1}{r} \,, \qquad \H(r) = \frac{\Lambda}{3} -r^2-2m\, r^3 \,, \label{sol:SchwAdS_ConfKundtCoord}
\end{equation}
which indicates that one should be careful about the dimensions of particular quantities, see footnote~\ref{footnote2}.

Finally, another useful parameterization of the spherical spacetimes, which removes coordinate singularities of the original metric (\ref{metric:SchwCoord}), belongs to the manifestly Robinson--Trautman class,
\begin{equation}
\dd s^2 = \Omega^2(\tilde r)\,(\dd \theta^2+\sin^2\theta\,\dd \phi^2)
-2\,\dd u\,\dd \tilde r+H(\tilde r)\,\dd u^2 \,. \label{metric:RT}
\end{equation}
which can be obtained from (\ref{metric:ConfKundt}) by
\begin{equation}
r = \int\!\!\frac{\dd \tilde r}{\Omega^2(\tilde r)}\, , \qquad H \equiv \Omega^{2}\, \H \,. \label{trans:ConfKundtToRT}
\end{equation}
The metric functions of the Schwarzschild--(anti-)de Sitter spacetime are
\begin{equation}
\Omega(\tilde{r})=\tilde{r} \,, \qquad H(\tilde{r})=-1+\frac{2m}{\tilde{r}} +\frac{\Lambda}{3}{\tilde{r}}^2\,. \label{sol:SchwAdS_RT}
\end{equation}

\subsection{Field equation within spherical symmetry}

The explicit form of the quadratic gravity field equations with non-vanishing $\Lambda$ and constant scalar curvature (\ref{eq:field_eq}), expressed using the conformal-to-Kundt parametrization (\ref{metric:ConfKundt}), was presented in \cite{SvarcPodolskyPravdaPravdova:2018} and in detail derived in \cite{PravdaPravdovaPodolskySvarc:2021}. The non-trivial constraints in the form of an autonomous system of two ordinary differential equations for the metric functions $\Omega(r)$ and $\H(r)$ reads
\begin{align}
\Omega\Omega''-2{\Omega'}^2 = & \ \tfrac{1}{3}k\, \B_1 \H^{-1} \,, \label{FieldEq1}\\
\Omega\Omega'{\H}'+3\Omega'^2{\H}+\Omega^2-\Lambda\Omega^4 = & \ \tfrac{1}{3}k \,\B_2 \,, \label{FieldEq2}
\end{align}
where the prime is the $r$-derivative and
\begin{align}
& \B_1 \equiv {\H}{\H}'''' \,, \label{DefB1}\\
& \B_2 \equiv {\H}'{\H}'''-\tfrac{1}{2}{{\H}''}^2 +2\,, \label{DefB2}
\end{align}
denote independent components of the Bach tensor. Moreover, it is also useful to employ the trace (\ref{eq:trace_general}), which becomes
\begin{equation}
{\H}\Omega''+{\H}'\Omega'+{\textstyle \frac{1}{6}} ({\H}''+2)\Omega = \tfrac{2}{3}\Lambda\Omega^3 \,. \label{eq:trace_explicit}
\end{equation}
Actually, the constraint~(\ref{FieldEq1}) follows from equations~(\ref{FieldEq2}) and (\ref{eq:trace_explicit}) and vice versa.

\section{Metric of the Bachian singularity\label{Sec:solution}}

 Using the conformal-to-Kundt coordinates (\ref{metric:ConfKundt}), a detailed discussion of generic solutions to (\ref{FieldEq1}) and (\ref{FieldEq2}) in the form of power series was given in \cite{PravdaPravdovaPodolskySvarc:2021}. However, analysis of specific subcases introduced by particular choices of the integration parameters and/or theory constants remained an open problem. The solution we present here belongs to such a peculiar fine-tuned class; however, its identification is more transparent starting from scratch. The obtained solution will be connected with the results of \cite{PravdaPravdovaPodolskySvarc:2021} in the following Section~\ref{Sec:Connection}.

\begin{figure}[h]
    \centering
    \includegraphics[width=0.45\textwidth]{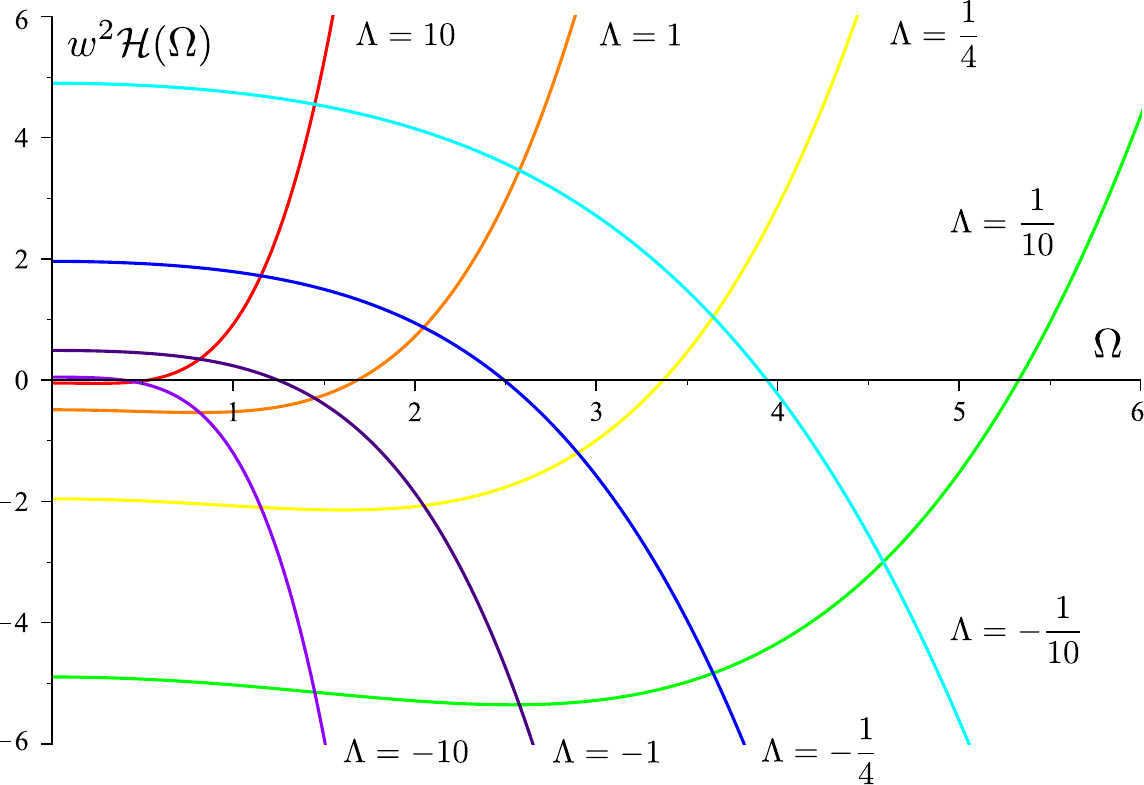}
    \caption{Dependence (\ref{H_sol_gauge}) of the function $\H$ scaled by the parameter $w^2$ on $\Omega$ for different values of $\Lambda$ related to the theory coupling constants by (\ref{cont_relation}). The signum of $\Lambda$ determines the staticity of the spacetime regions separated by the horizon $\H=0$ via condition (\ref{eq:KillingVec}).}\label{fig:H_Omega_cK}
\end{figure}

Let us assume the metric function $\Omega$ and $\H$ in a simple polynomial form,
\begin{align}
\Omega(r) =& q_1r+q_0\,, \label{Omega_ansatz}\\
\H(r) =& s_4r^4+ s_3r^3+s_2r^2+s_1r+s_0\,, \label{H_ansatz}
\end{align}
with $q_i$ and $s_i$ being constants. Then the field equation (\ref{FieldEq1}) implies
\begin{equation}
s_4=-\frac{q_1^2}{4k} \,.
\end{equation}
Subsequently, the constraints following from particular orders in $r$ of equation (\ref{FieldEq2}) read
\begin{align}
\Lambda &=  -\frac{9}{4k} \,, \label{Lambda_condition} \\
s_3 &=  -\frac{q_0q_1}{k} \,, \\
s_2 &=  -\frac{1}{7}-\frac{3q_0^2}{2k} \,, \\
s_1 &=  -\frac{q_0}{q_1}\left(\frac{2}{7}+\frac{q_0^2}{k}\right) , \\
s_0 &=  \frac{1}{q_1^2}\left(\frac{32k}{147}-\frac{q_0^2}{7}-\frac{q_0^4}{4k}\right) ,
\end{align}
where (\ref{Lambda_condition}) represents the fine tuning of the theory constants. The solution is still subject to the gauge transformation~(\ref{GaugeTrans}). Its application together with the inverse of (\ref{Lambda_condition}) in general implies\footnote{\label{footnote2}Note that the parameter $w$ is not dimensionless since $\Omega$ has a dimension of length, however, the dimension of $r$ itself is $1/\mbox{length}$.}
\begin{align}
\Omega(r) &= wr+v\,, \label{Omega_sol_gauge}\\
\H(r) &= \frac{1}{w^2}\left(\frac{\Lambda}{9}\,\Omega^4-\frac{1}{7}\,\Omega^2-\frac{24}{49\Lambda}\right) , \label{H_sol_gauge}
\end{align}
where we have denoted the coefficients in (\ref{Omega_sol_gauge}) in the same way as they are introduced in (\ref{Omega_ansatz}), however, now the $r$-coordinate gauge freedom is explicitly included, namely
\begin{equation}
w\equiv \lambda^{-1} q_1 \,, \qquad v\equiv q_1\nu+q_0 \,.
\end{equation}
This shows that there are no free physical parameters arising from the integration of the field equations, and the only freedom is hidden in the constant $\Lambda$ related to the theory couplings as
\begin{equation}
\Lambda=-\frac{9\,\gamma}{4(\alpha+18\beta)} \,, \label{cont_relation}
\end{equation}
where (\ref{k_definition}) and (\ref{Lambda_condition}) were used. The typical dependence of the metric function $\H$  (scaled by $w^2$) given by (\ref{H_sol_gauge}) on the second metric function $\Omega$ for several values of the parameter~$\Lambda$ is visualized in Figure~\ref{fig:H_Omega_cK}.

\subsection{Role of the metric functions\label{SubSec:MetricFunctions}}

In the metric form (\ref{metric:ConfKundt}), the function $\Omega(r)$ encodes the areal radius of the sphere $r=\mbox{const}$, $u=\mbox{const}$, since the area of such a surface is $4\pi\Omega^2$, while the function $\mathcal{H}$ gives the horizon location by the condition $\H(\Omega_h)=0$, which for (\ref{H_sol_gauge}) explicitly becomes
\begin{equation}
\Omega_h^2 = \frac{3}{ 14\Lambda} \left(3 + \sgnL \sqrt{105} \right) . \label{horizon_cK_coord_Omega}
\end{equation}
The existence of a horizon is implied by the causal behavior of the Killing vector field $\xi=\partial_{t}$ in terms of the classic parametrization (\ref{metric:SchwCoord}). Using the conformal-to-Kundt coordinates (\ref{metric:ConfKundt}), it takes the form
\begin{equation}
	\mathbf{\xi} = \partial_{u} \,, \quad \mbox{with} \qquad \mathbf{\xi} \cdot \mathbf{\xi} = \Omega^{2} \H  \,, \label{eq:KillingVec}
\end{equation}
and becomes null if $\mathcal{H}=0$. The surface area of such a horizon then simply becomes
\begin{equation}
    \mathcal{A} = 4 \pi \Omega_h^{2} \,. \label{horizon:area}
\end{equation}
Based on the norm (\ref{eq:KillingVec}), the spacetime is static for $\H<0$, where $\xi$ is time-like. For $\Lambda<0$ it corresponds to the region with $\Omega>\Omega_h$ and vice versa; see (\ref{H_sol_gauge}) and figure~\ref{fig:H_Omega_cK}.

To determine the explicit value $r_h$ of $r$-coordinate that satisfies the horizon condition $\H(r_h)=0$ we simply combine (\ref{Omega_sol_gauge}) and (\ref{horizon_cK_coord_Omega}) and express $r_h$ through $\Omega_h=\Omega(r_h)$ as $r_h=w^{-1}\left(\Omega_h-v\right)$, namely
\begin{align}
\rh &= \frac{1}{w}\left\{\pm\left[\frac{3}{ 14\Lambda} \left(3 + \sgnL \sqrt{105} \right)\right]^{1/2}-v\right\} . \label{cK_r_hozinot_explicit}
\end{align}
Here, we still keep the gauge parameters fully generic. In Subsection~\ref{SubSec:NaturalGauge}, we show their natural choice; however, in Section~\ref{Sec:Connection} revisiting the previous results, another (less natural) gauge fixing will necessarily be applied.

\subsection{Curvature}

In the conformal-to-Kundt form (\ref{metric:ConfKundt}) of the spherical spacetimes, the coordinate $u$ labels null hypersurfaces that foliate the spacetime manifold, while $r$ is an affine parameter along the geodesics that generate such a foliation. The remaining coordinates $\theta$ and $\phi$ cover the spheres determined by constant values of $u$ and $r$. The null hypersurface tangent (and normal) vector field can thus be written as
\begin{equation}
\vect{k}=\Omega^{-2}\,\partial_{r} \,,
\end{equation}
which can be supplemented by 
\begin{equation}
\vect{l}=\frac{1}{2}\H\partial_r+\partial_u \,, \qquad \vect{m}=\frac{1}{\sqrt{2}\Omega}\left(\partial_{\theta}-\frac{i}{\sin\theta} \partial_{\phi}\right),
\end{equation}
to obtain null frame $\{\boldk,\,\boldl,\,\boldm,\,\bar{\boldm}\}$ satisfying
\begin{equation}
\vect{k}\cdot\vect{l}=-1\,, \qquad \vect{m}\cdot\bar{\vect{m}}=1 \,.
\end{equation}
The expansion of the null foliation generator $\vect{k}$ is
\begin{equation}
\rho=-\frac{\Omega_{,r}}{\Omega^3}=-\frac{w}{\Omega^3} \,,
\end{equation}
which provides the interpretation of $w$ as a scaling of expansion of $\boldk$ taken on a given spherical surface with the area $4\pi\Omega^2$. 
Note that for the pure Schwarzschild--(anti-)de Sitter solution (\ref{sol:SchwAdS_ConfKundtCoord}), where the gauge (\ref{GaugeTrans}) is already trivially fixed, we get $\rho= -r^{-1}$.

With respect to the frame above, the Weyl tensor is manifestly of algebraic type D, and its only nontrivial component becomes
\begin{equation}
\Psi_2= -\frac{\Lambda}{9} - \frac{1}{7\Omega^{2}} \,, \label{Psi2_Omega_cK}
\end{equation}
which can be again compared with the Schwarzschild--(anti-)de Sitter case (\ref{sol:SchwAdS_ConfKundtCoord}) leading to $\Psi_2= -m\,r^{-3}$. For $\Omega(r)\rightarrow\infty$ the Weyl scalar (\ref{Psi2_Omega_cK}) becomes a non-vanishing constant, which implies that the spacetime asymptotic behavior is not conformally flat.

The nontrivial components of the Ricci tensor \cite{Stephanietal:2003} are
\begin{align}
\Phi_{00} =&\, \frac{2w^2}{\Omega^6} \,, \nonumber \\
\Phi_{11} =&\, -\frac{2\Lambda}{9} + \frac{2}{7\Omega^{2}} - \frac{12}{49\Lambda\Omega^{4}} \,, \nonumber \\
\Phi_{22} =&\, \frac{1}{w^2}\left(\frac{\Lambda^2}{162}\Omega^6 - \frac{\Lambda}{63}\Omega^{4} \right. \nonumber\\
&\hspace{10.0mm} \left. - \frac{13}{294}\Omega^{2} + \frac{24}{343\Lambda} + \frac{288}{2401\Lambda^2\Omega^{2}}\right), \label{Ricci_components_Omega_cK}
\end{align}
which demonstrates its algebraic generality \cite{Ortaggioetal:2012} and highlights the difference in comparison with GR, where, due to the vacuum field equations, all the Ricci components are identically zero. Similarly, these components correspond to the nontrivial Bach tensor through the field equations (\ref{eq:field_eq}). The explicit form of the projections (\ref{Psi2_Omega_cK}) and (\ref{Ricci_components_Omega_cK}) indicates the presence of a curvature singularity at $\Omega(r)=0$. This can be directly observed from the Kretschmann scalar $K\equiv R_{abcd}R^{abcd}$, namely
\begin{align}
K=& \ \frac{136\Lambda^2}{27} - \frac{64\Lambda}{21\Omega^{2}} \nonumber\\ 
& \ +\frac{832}{147\Omega^{4}} - \frac{768}{343\Lambda\Omega^6}+ \frac{13824}{2401\Lambda^2\Omega^{8}} \,. \label{Riemann_sqr_generic_Omega_cK}
\end{align}
It contains contributions of the constant scalar curvature $R=4\Lambda$, square of the Weyl tensor,
\begin{equation}
C_{abcd}C^{abcd}= \frac{16\Lambda^2}{27} + \frac{32\Lambda}{21\Omega^{2}} + \frac{48}{49\Omega^{4}} \,, \label{Weyl_sqr_generic_Omega_cK}
\end{equation}
and the square of the Ricci tensor,
\begin{align}
R_{ab}R^{ab}= & \  \frac{44\Lambda^2}{9}- \frac{16\Lambda}{7\Omega^{2}} \nonumber \\
& \  +\frac{344}{147\Omega^{4}}- \frac{384}{343\Lambda\Omega^6} + \frac{6912}{2401\Lambda^2\Omega^{8}} \,. \label{Ricci_sqr_generic_Omega_cK}
\end{align}
Comparison of the expressions (\ref{Riemann_sqr_generic_Omega_cK}), (\ref{Weyl_sqr_generic_Omega_cK}), and (\ref{Ricci_sqr_generic_Omega_cK}) shows that the dominant singular contribution at ${\Omega(r)=0}$ arises from the Ricci part connected to the Bach tensor by the gravitational law (\ref{eq:field_eq}). Vice versa, at infinity all these scalars are constant, in particular, the Kretschmann scalar becomes $K({\Omega(r)}\rightarrow\infty)=\frac{136\Lambda^2}{27}$. Such a behavior of the curvature components in the asymptotic region can be directly connected with tidal deformations of the congruence of freely falling observers; see the last part of Section~\ref{Sec:GeodDev}.

Finally, the combination of these results with those on staticity obtained in Section~\ref{SubSec:MetricFunctions} gives the picture that the case of negative cosmological parameter ($\Lambda<0$) corresponds to the singularity at $\Omega=0$ surrounded by the dynamical region hidden below the black-hole horizon at $\Omega_h$ and followed by the outer static region, whereas the positive cosmological parameter ($\Lambda>0$) leads to the naked singularity, the inner static region bordered by the cosmological horizon, and the dynamical region outside.

\subsection{Natural gauge choice\label{SubSec:NaturalGauge}}

Various particular choices of the gauge parameters $\lambda$ and $\nu$ can be conveniently applied. For example, it is natural to partially employ this freedom and set
\begin{equation}
v=0 \,, \quad \mbox{i.e.,}\quad \nu=-\frac{q_0}{q_1} \,, \label{nu_gauge}
\end{equation}
which using (\ref{trans:SchwToConfKundt}) guaranties the correspondence
\begin{equation}
\Omega(r)=0 \quad \leftrightarrow \quad r=0 \quad \leftrightarrow \quad \bar{r}=0 \,.
\end{equation}
Then, the metric functions (\ref{Omega_sol_gauge}) and (\ref{H_sol_gauge}) of the conformal-to-Kundt line element (\ref{metric:ConfKundt}) become
\begin{align}
\Omega(r) &= w\,r \,, \label{Omega_sol} \\
\H(r) &= \frac{\Lambda}{9}w^2\,r^4-\frac{1}{7}\,r^2-\frac{24}{49\Lambda}w^{-2}\,\,, \label{H_sol}
\end{align}
where $w$ remains an unspecified gauge parameter allowing us to fix the geometric meaning and dimension of the `radial' coordinate $r$. For example, the value of $w$ can be trivially set to $w=1$ due to $\lambda=q_1$. Then the value of the coordinate~$r$ has the direct meaning of the standard surface radius, since the area of a sphere with $r=\mbox{const}$, $u=\mbox{const}$ becomes $4\pi r^2$. However, to simplify particular expressions, other choices may be more convenient; see, e.g., Robinson--Trautman form of the metric functions in Subsection~\ref{SubSec:AltCoord}.

\subsection{Thermodynamical quantities}

The temperature $T$ of the (black-hole) horizon located at ${r=r_h}$ is given as a rescaling of its surface gravity $\kappa$,
\begin{align}
    T = \frac{\kappa}{2 \pi} \,,
\end{align}
where $\kappa$ is defined as
\begin{equation}
    \kappa^{2} \equiv -\frac{1}{2}\left(\nabla_{a}\xi_{b}\right)\left(\nabla^{a} \xi^{b}\right)\big|_{r=r_{h}} \,,
\end{equation}
with $\xi^{a}$ being the Killing vector field (\ref{eq:KillingVec}). The explicit calculation of the surface gravity for the metric (\ref{metric:ConfKundt}) with (\ref{Omega_sol_gauge}) and (\ref{H_sol_gauge}), and its evaluation at $r_h$, results in
\begin{align}
    \kappa &= -\frac{1}{2}\mathcal{H}'\big|_{r=r_{h}}
    = -\frac{\Omega_h}{w}\left(\frac{2\Lambda}{9}\Omega_h^2 - \frac{1}{7}\right) , \label{eq:surface_grav}
\end{align}
where $\Omega_h$ is given by (\ref{horizon_cK_coord_Omega}).

The black-hole entropy depends not only on the spacetime metric but also on the theory itself. Employing the Wald definition \cite{Wald:1993, IyerWald:1994} we have
\begin{equation}
S=\frac{2\pi}{\kappa}\,\oint \mathbf{Q}\,, \label{enthropy_def:Wald}
\end{equation}
where $\mathbf{Q}$ is the Noether-charge 2-form that must be evaluated on the horizon $r_h$. In general, for the conformal-to-Kundt spherical metric (\ref{metric:ConfKundt}) within quadratic gravity~(\ref{eq:quadratic_gravity_action}) such 2-form becomes
\begin{align}
\mathbf{Q} = &-\frac{\Omega^2\, \H'}{16\pi} \left[\gamma +\frac{4}{3}\Lambda(\alpha+6\beta) +\frac{4}{3}k\alpha\,\frac{\B_1+\B_2}{\Omega^4}\right]\bigg|_{r=r_{h}} \nonumber \\
& \times \, \sin\theta\,\dd\theta\wedge\dd\phi \,, \label{enthropy:Q}
\end{align}
see \cite{SvarcPodolskyPravdaPravdova:2018, PravdaPravdovaPodolskySvarc:2021} for more details. Using the Bach component definitions (\ref{DefB1}) and (\ref{DefB2}) evaluated at $r_h$ for the explicit solution (\ref{Omega_sol_gauge}) and (\ref{H_sol_gauge}) with the constraint (\ref{Lambda_condition}), it further simplifies, namely,
\begin{align}
\mathbf{Q} = &\frac{\kappa\,\Omega_h^2}{8\pi} \left[\gamma +\frac{4}{3}\Lambda(\alpha+6\beta) -\frac{480}{49} \frac{\alpha}{\Lambda\Omega_h^4}\right] \sin\theta\,\dd\theta\wedge\dd\phi \,. \label{enthropy:Q:spec}
\end{align}
Moreover, the parameter $\Lambda$ is related to the coupling constants $\alpha$, $\beta$, and $\gamma$ via (\ref{cont_relation}) and $\Omega_h$ is given in terms of $\Lambda$ by (\ref{horizon_cK_coord_Omega}) and determines the horizon area as (\ref{horizon:area}). Finally, this allows us to express the entropy (\ref{enthropy_def:Wald}) with (\ref{enthropy:Q:spec}) as
\begin{align}
\!S &= \frac{\gamma}{4}\mathcal{A}\left[1 -3\frac{\alpha + 6 \beta}{\alpha + 18 \beta} + \frac{5\left(3-\sgnL\sqrt{105}\right)^2\alpha}{96(\alpha + 18 \beta)}\right]\!.
\end{align}
This expression allows for a comparison with GR, namely, the first term is a standard GR result, the second one corresponds to the quadratic gravity correction to the Schwarzschild--(A)dS solution with specific $\Lambda$ given by~(\ref{cont_relation}), while the third term is a specific geometric contribution arising from the non-trivial Bach tensor components. 

\subsection{Geodesics}

The purpose of this section is to describe the motion of an observer along a geodesic $\gamma(\tau)$ with four-velocity ${\boldu=\dot{r}\,\partial_r+\dot{u}\,\partial_u+\dot{\theta}\,\partial_\theta+\dot{\phi}\,\partial_\phi}$ in spacetime (\ref{metric:ConfKundt}). In general, the components of the geodesic equation $\nabla_{_{\boldu}}\boldu=0$ become
\begin{align}
\ddot{r} = &  -2\omega\,\dot{r}^2+\left(2\omega\H+\H'\right)\dot{r}\dot{u} -\left(\frac{1}{2}\H'+\omega\H\right)\!\H\,\dot{u}^2 \nonumber \\
&-\omega\H\left(\dot{\theta}^2 + \sin^2\theta\,\dot{\phi}^2\right) \,, \label{eq:geod2_r}\\
\ddot{u} = &  -\left(\frac{1}{2}\H'+\omega\H\right)\dot{u}^2-\omega \left(\dot{\theta}^2 + \sin^2\theta\,\dot{\phi}^2 \right) \,, \label{eq:geod2_u}\\
\ddot{\theta} = &  -2\omega\,\dot{r}\dot{\theta} +\sin\theta\cos\theta\,\dot{\phi}^2 \,, \label{eq:geod2_theta}\\
\ddot{\phi} = &  -2\left(\omega\,\dot{r}+\cot\theta\,\dot{\theta} \right)\dot{\phi}\,, \label{eq:geod2_phi}
\end{align}
where $\omega\equiv\Omega'\Omega^{-1}$, prime is the $r$-derivative, dot denotes the derivative with respect to the affine parameter $\tau$, and the functions $\Omega$ and $\H$ must be substituted from (\ref{Omega_sol_gauge}) and (\ref{H_sol_gauge}). To specify a particular geodesic observer in spherical geometries (\ref{metric:ConfKundt}), symmetries can advantageously be employed. Namely, in addition to the metric itself, the spacetime admits the Killing tensor
\begin{align}
    \mathbf{K} &= \Omega^4\left( \dd\theta\,\dd\theta +\sin^2\theta\, \dd\phi\,\dd\phi\right),
\end{align}
and the Killing vectors $\mathbf{\xi} = \partial_{u}$ and $\mathbf{\eta} = \partial_{\phi}$. This allows for the definition of the conserved quantities as
\begin{align}
    \sigma &= {u}^a\,g_{ab}\, {u}^b \,, &
    E &= -{u}^a {\xi}_a \,, \nonumber \\
    L^2 &= {u}^a \,{K}_{ab}\,{u}^b \,, &
    L_z &= {u}^a {\eta}_a \,, \label{conserved_quantities}
\end{align}
where ${\sigma=0,\,\pm1}$ is the 4-velocity normalization and ${L^2\geq0}$. Then, the constraints on the motion along $\gamma({\tau)}$ can be written in the form of the first ODE system
\begin{align}
    \dot{r}^2 &=\frac{1}{\Omega^4}\left[E^2-\left(\sigma\Omega^2-L^2\right)\mathcal{H}\right], \label{geod_eq_dot_r} \\ 
    \dot{u} &= \frac{1}{\Omega^2\mathcal{H}}\left(-E+\Omega^2\dot{r}\right), \label{geod_eq_dot_u} \\
    \dot{\theta}^2 &=\frac{1}{\Omega^4}\left(L^2-\frac{L_z^2}{\sin^2\theta}\right), \label{geod_eq_dot_theta} \\ 
    \dot{\phi} &=\frac{L_z}{\Omega^2\sin^2\theta}\,. \label{geod_eq_dot_phi}
\end{align}
To specify the initial data for the original geodesic equations (\ref{eq:geod2_r})--(\ref{eq:geod2_phi}), we need to prescribe the initial position $\{r_0,\, u_0,\, \theta_0,\, \phi_0\}$ and the corresponding four-velocity. These initial conditions are related to the constants of motion (\ref{conserved_quantities}) through conditions (\ref{geod_eq_dot_r})--(\ref{geod_eq_dot_phi}). However, admitted choices are restricted by the system itself, that~is,
\begin{align}
    E^2&\geq\mathcal{H}(r_0)\left(\sigma\Omega^2(r_0)-L^2\right)\,,&
    L^2&\geq \frac{L_z^2}{\sin^2\theta_0} \,, \label{eq:E_L_constraints}
\end{align}
where the right-hand side of the first inequality corresponds to value of the effective potential for radial motion at given ${r=r_0}$. Actually, equations~(\ref{geod_eq_dot_r}) and (\ref{geod_eq_dot_theta}) restrict the possible ranges of radial and azimuthal coordinates where a particle with given energy and angular momentum can be located. Due to spherical symmetry, the motion of a particle is effectively planar, and thus we can consider the motion in the equatorial plane (${\theta=\pi/2}$ can be fixed for an individual particle without loss of generality).

An interesting class of trajectories corresponds to circular orbits, i.e. orbits with vanishing velocity and acceleration in radial direction, namely $\dot{r}=0=\ddot{r}$. The relevant conditions for a specific compatible value of the coordinate $r$ read
\begin{align}
E^2 = & \, -\left(L^2+\Omega^2\right)\mathcal{H} \,, \label{eq:co_general_E}\\
L^2\H' = & \, \sigma\Omega\left(\Omega\mathcal{H}'+2\mathcal{H}\Omega'\right). \label{eq:co_general_L}
\end{align}

\begin{figure}[h]
    \centering
    \includegraphics[width=0.45\textwidth]{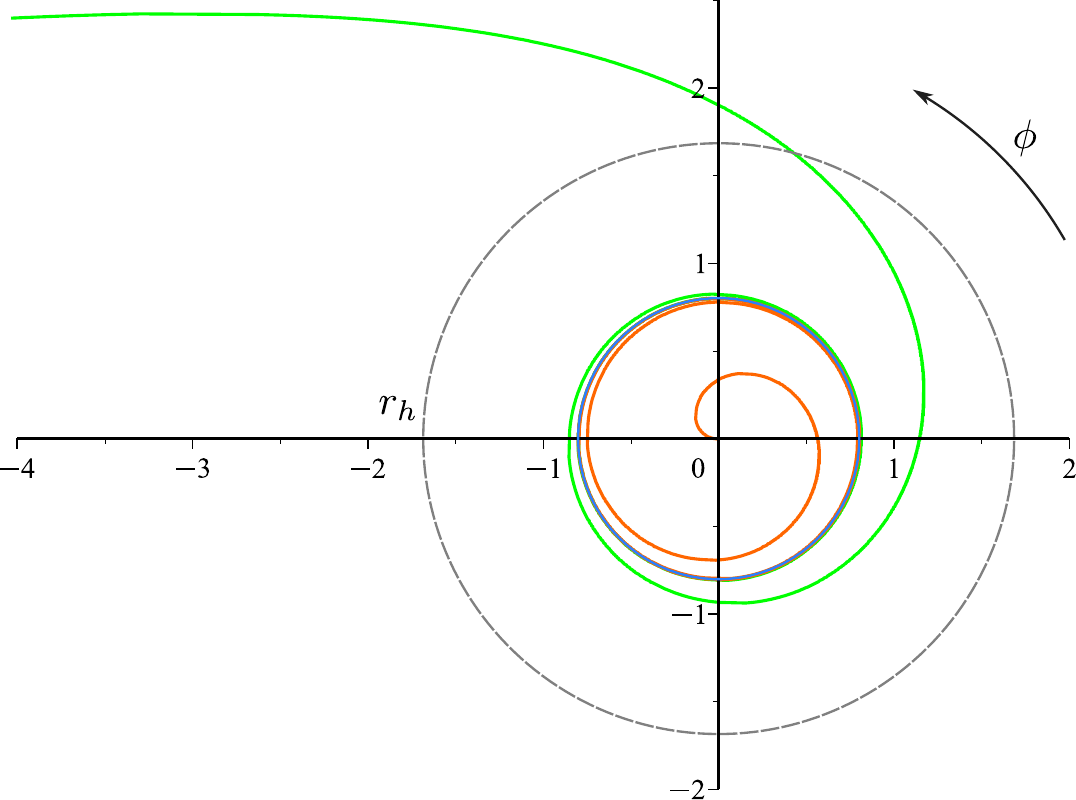}
    \caption{For $\Lambda=1$, we plot the circular photon orbit (blue -- partially covered by nearby non-circular red and green trajectories) at ${r_{\mathrm{po}}=3/\sqrt{14}}\approx0.8$ and a pair of typical nearby light rays starting with ${\dot{r}(0)=0}$ at ${r_0=r_{\mathrm{po}}\pm10^{-9}}$. The inner one (red) spirals into the central singularity, while the outer one (green) radially escapes due to $\Lambda>0$. The motion in the equatorial plane $\theta=\pi/2$ is considered. The cosmological horizon $r_h$ is plotted as a grey dashed line.}\label{fig:photon_orbit}
\end{figure}

Interestingly, for light rays ($\sigma=0$), there exists a circular photon orbit $r(t)=r_\mathrm{po}=\mbox{const}$, only for $\Lambda>0$. Based on the constraint (\ref{eq:co_general_L}), the $r_\mathrm{po}$ value is given by ${\H'(r_{\mathrm{po}})=0}$, which using (\ref{Omega_sol_gauge}) and (\ref{H_sol_gauge}) with the simple gauge choice $w=1$ and $v=0$ gives ${r_{\mathrm{po}}^2=9/({14\Lambda})}$. For the typical unstable behavior of the nearby trajectories above/below the photon orbit, see Figure~\ref{fig:photon_orbit}.

\begin{figure}[h]
    \centering
    \includegraphics[width=0.45\textwidth]{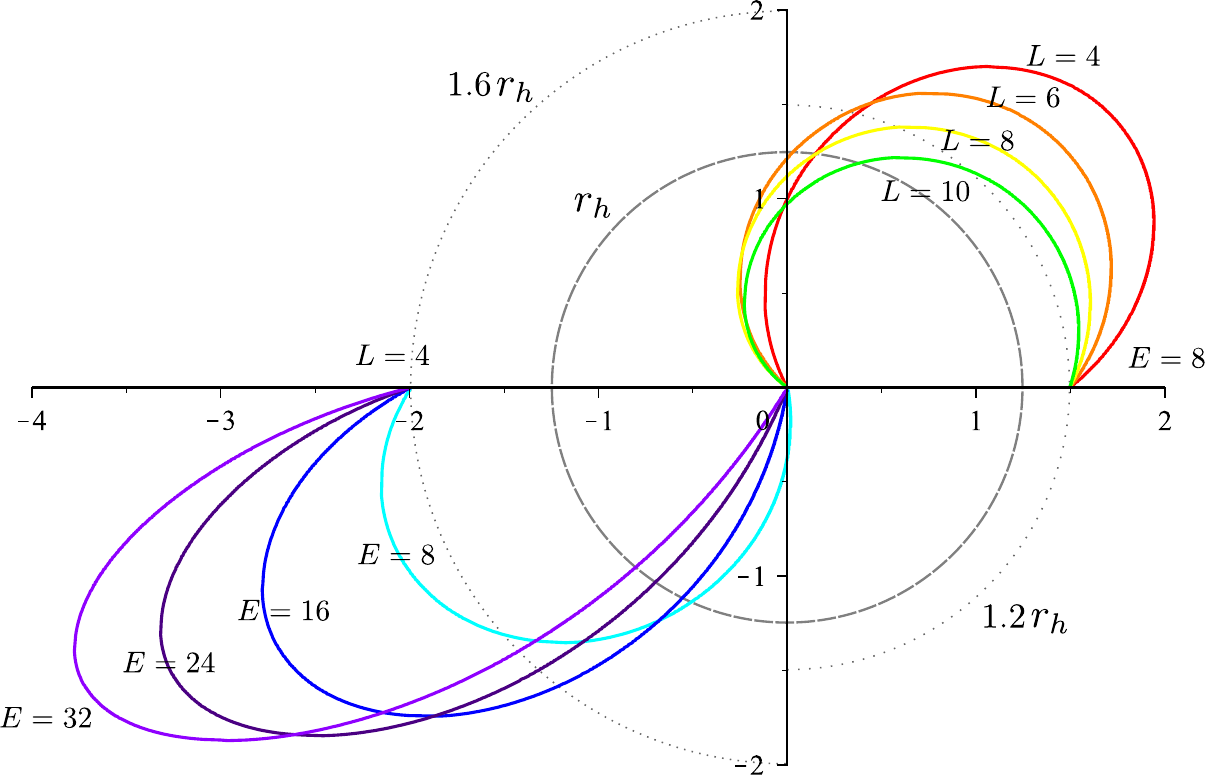}
    \caption{We visualize typical timelike ($\sigma=-1$) trajectories for different values of the parameters $E$ and $L_z=L$ (so that the particles remain in equatorial plane which is allowed due to the spherical symmetry) in the case with $\Lambda=-1$. The two quadruplets of test particles start at two different radii (indicated by dotted semicircles), and all trajectories end up in the central singularity. Particular values of the initial data are plotted in the figure. The maximal value of the radial coordinate is restricted by (\ref{eq:E_L_constraints}). The black-hole horizon $r_h$ is plotted as a dashed line.}\label{fig:timelike_Lambda_m}
\end{figure}

\begin{figure}[h]
    \centering
    \includegraphics[width=0.45\textwidth]{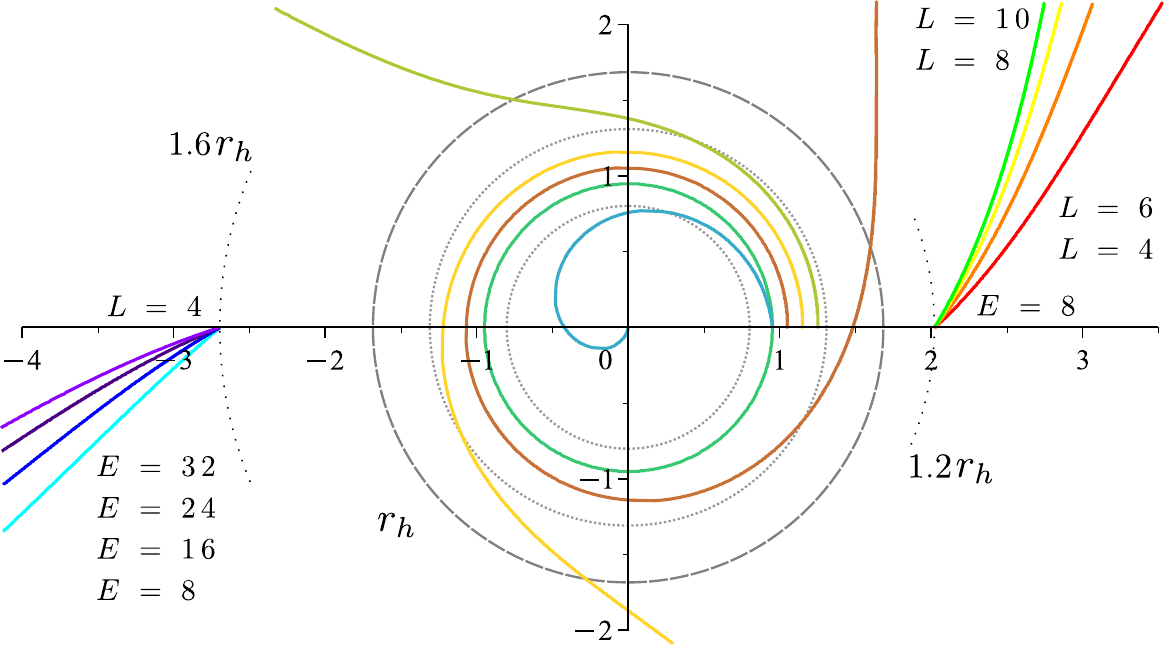}
    \caption{For $\Lambda=1$ we plot typical timelike ($\sigma=-1$) trajectories for different values of the parameters $E$ and $L_z=L$.  The cosmological horizon at $r=r_h$ is plotted as a dashed line. The two quadruplets of test particles outside the horizon start at two different radii (indicated by dotted arcs), and all trajectories escape the centre. Particular values of the initial data are plotted in the figure. Below the horizon pair of dotted circles represents radii (\ref{circular_boundaries}) forming the boundary of a region admitting circular time orbits, where the data are given by (\ref{eq:co_general_E}) and (\ref{eq:co_general_L}). One such orbit is plotted in green. The remaining trajectories slightly violate the data constraint for circular orbits. In particular, the inner curve ending in the singularity has $L$ less than the circular value (\ref{eq:co_general_L}), while three trajectories leaving the central region have $E$ greater than~(\ref{eq:co_general_E}).} \label{fig:timelike_Lambda_p}
\end{figure}

In general, for timelike trajectories ($\sigma=-1$), the parameter $L$ determines the total angular momentum of the particle, while $L_z$ is the momentum with respect to the axis. For the equatorial trajectories $L=L_z$. Since $\H$ is a negative decreasing function above the horizon $r_h$ for $\Lambda<0$, see Figure~\ref{fig:H_Omega_cK}, the first constraint in (\ref{eq:E_L_constraints}) implies that there is a maximum distance $r_{\mathrm{max}}>r_h$ which can be reached by a timelike particle with a given energy~$E$; as an illustration, see Figure~\ref{fig:timelike_Lambda_m}.
The timelike circular geodesics may exist only for $\Lambda>0$ and are located below the cosmological horizon. The range of admissible radii (using the simplest gauge $v=0$ and $w=1$) is
\begin{align}
    r&\in\left(\frac{3}{\sqrt{14\Lambda}}\,,\ \frac{2\sqrt{3}}{\sqrt{7\Lambda}}\right) , \label{circular_boundaries}
\end{align}
and constants of motion are given by (\ref{eq:co_general_E}) and (\ref{eq:co_general_L}). At the inner rim, the particle has a maximum velocity limiting to the circular photon orbit discussed above, while at the outer rim it has vanishing angular momentum $L^2=0$ and therefore resides at a fixed spatial point.  In fact, these geodesics are unstable maxima of the effective potential. For the typical behavior of timelike geodesics with $\Lambda>0$ see Figure~\ref{fig:timelike_Lambda_p}.

\subsection{Geodesic deviation\label{Sec:GeodDev}}

This part is a direct application of results obtained in Section~V.C.3 of \cite{PravdaPravdovaPodolskySvarc:2021}. We aim to describe the tidal effects, in principle, measurable by the radially falling test observer. This shows that the contribution of the Bach tensor~(\ref{DefB1}) and (\ref{DefB2}) can be, in principle, identified by the geodesic deviation. However, here, we want to focus mainly on the dominant curvature contribution to relative motion of nearby observers in regions far from the central singularity ($\Omega\rightarrow\infty$) to distinguish the anisotropy in comparison with the maximally symmetric (anti-)de Sitter backgrounds.

In particular, as a projection of the classic equation of geodesic deviation with the separation vector~$\boldZ$ onto an orthonormal frame ${\{\bolde_{(0)}, \bolde_{(1)}, \bolde_{(2)}, \bolde_{(3)}}\}$ normalized as ${\bolde_{(a)}\cdot\bolde_{(b)}=\eta_{ab}}$, associated with the observer in such a way that the time-like vector ${\bolde_{(0)}}$ is its $4$-velocity ${\boldu}$, we get
\begin{equation}
\ddot Z^{(\rm{i})}= R^{(\rm{i})}_{\quad(0)(0)(\rm{j})}\,Z^{(\rm{j})} \,,\qquad
\rm{i},\,\rm{j}=1,\,2,\,3\,, \label{GeodDev:projection}
\end{equation}
with
${\ddot Z^{(\rm{i})}\equiv e^{(\rm{i})}_a\,{Z^a}_{;cd}\, u^c u^d}$ denoting the relative accelerations of the freely falling nearby test particles and ${R_{(\rm{i})(0)(0)(\rm{j})}= R_{abcd} \,e^a_{(\rm{i})}u^b u^c e^d_{(\rm{j})}}$ describing their source given by the spacetime curvature. To reveal particular effects of the gravitational field encoded in (\ref{GeodDev:projection}), it is convenient to decompose the Riemann tensor into the Weyl tensor, the Ricci tensor, and the Ricci scalar, substitute for the Ricci curvature from the field equations~(\ref{eq:field_eq}), and simultaneously express orthonormal components of the  Weyl tensor in terms of the Newman--Penrose null frame scalars. The natural relation between the orthonormal and null frame ${\{\mathbf{k}, \mathbf{l}, \mathbf{m} \}}$ is
\begin{align}
&\mathbf{k}=\frac{1}{\sqrt{2}}(\boldu+\bolde_{(1)})\,, \qquad \mathbf{l}=\frac{1}{\sqrt{2}}(\boldu-\bolde_{(1)})\,, \nonumber \\
&\mathbf{m}=\frac{1}{\sqrt{2}}\left(\bolde_{(2)}-i\bolde_{(3)}\right) , \label{frame:null}
\end{align}
see \cite{PravdaPravdovaPodolskySvarc:2021, PodolskySvarc:2012}. Since we are interested in radially falling observers in the spacetime (\ref{metric:ConfKundt}) with (\ref{Omega_sol_gauge}) and (\ref{H_sol_gauge}), we set ${\dot{\theta}=0=\dot{\phi}}$ and the orthonormal frame can be chosen~as
\begin{equation}
\begin{aligned}
& \bolde_{(0)}=\boldu= \dot{r}\,\partial_r +\dot{u}\,\partial_u \,, \\
& \bolde_{(1)}= \frac{1}{2}\left[( {\Omega^2\dot{u}} )^{-1}-{\H}\dot{u}\right]\partial_r -\dot{u}\,\partial_u \,, \\
& \bolde_{(2)}= \Omega^{-1}\,\partial_{\theta} \,, \\ 
& \bolde_{(3)}= \left(\Omega\sin\theta\right)^{-1}\partial_{\phi} \,,
\end{aligned}
\label{frame:orthonormal}
\end{equation}
where ${\dot{r}=\frac{1}{2}\left[({\Omega^2\,\dot{u}})^{-1}+{\H}\dot{u}\right]}$ through the 4-velocity normalization ${\boldu\cdot\boldu=-1}$, i.e. (\ref{conserved_quantities}) with $\sigma=-1$. Finally, the projected equation of geodesic deviation (\ref{GeodDev:projection}) with (\ref{frame:null}) and (\ref{frame:orthonormal}) becomes
\begin{align}
\ddot{Z}^{(1)} = &
\frac{\Lambda}{3}\,Z^{(1)} -2\Psi_{2}\,Z^{(1)} \nonumber\\
& +2k\left(B_{(1)(1)}-B_{(0)(0)}\right)Z^{(1)}, \label{InvGeoDevFinal1}\\
\ddot{Z}^{(i)} = &\frac{\Lambda}{3}\,Z^{(i)} + \Psi_{2}\,Z^{(i)}  \nonumber\\
& +2k\left(B_{(i)(j)} \,Z^{(j)}-B_{(0)(0)}\,Z^{(i)}\right), \label{InvGeoDevFinal2}
\end{align}
where the range of the indices is ${i,\,j=2,\,3}$. The general deformation thus consists of the isotropic motion caused directly by the term $\Lambda$, the classical Newtonian tidal deformation $\Psi_{2}$ given explicitly as ${\Psi_{2}=-\frac{1}{12}\,\Omega^{-2}({\H}''+2)}$ and exactly corresponding to (\ref{Psi2_Omega_cK}), since $\mathbf{k}=\dot{u}^{-1}\,\boldk$, $\mathbf{l}=\dot{u}\,\boldl$, and $\mathbf{m}=\boldm$, and the quadratic gravity corrections encoded in the orthonormal components of the Bach tensor. Its relevant orthonormal projections are
\begin{align}
B_{(0)(0)}&= \frac{-1}{24\,\Omega^6\dot{u}^2} \left[\!\left(1-\Omega^2{\H}\dot{u}^2\right)^2\,\frac{\B_1}{\H} -2\Omega^2\dot{u}^2\B_2\right]\!,\! \\
B_{(1)(1)}&= \frac{-1}{24\,\Omega^6\dot{u}^2} \left[\!\left(1+\Omega^2{\H}\dot{u}^2\right)^2\,\frac{\B_1}{\H} +2\Omega^2\dot{u}^2\B_2\right]\!,\! \\
B_{(0)(1)}&= \frac{-1}{24\,\Omega^6\dot{u}^2} \left(1-\Omega^4{\H}^2\dot{u}^4\right)\frac{\B_1}{\H} \,, \\
B_{(i)(j)}&= \frac{\delta_{ij}}{12\,\Omega^4}\left(\B_1+\B_2\right) \,,
\end{align}
where $\B_1$ and $\B_2$ are given by~(\ref{DefB1}) and (\ref{DefB2}), and the components $i,\,j$ are proportional to the $\theta$ and $\phi$ directions, respectively. Combining all the expressions\sk{,} we get
\begin{align}
\ddot{Z}^{(1)} =&
\frac{\Lambda}{3}\,Z^{(1)} + \, \frac{1}{6}\, \Omega^{-2}\big({\H}''+2\big)\,Z^{(1)}\, \nonumber \\
& -\,\frac{1}{3}\,k\,\Omega^{-4}\left(\B_1+\B_2\right)Z^{(1)} \,, \label{InvGeoDevBH1}\\
\ddot{Z}^{(i)} =& \frac{\Lambda}{3}\,Z^{(i)} - \frac{1}{12}\,\Omega^{-2}\left({\H}''+2\right)Z^{(i)} \nonumber \\
& +\frac{1}{12}\,k\,\Omega^{-4}\left((\Omega^2\H\dot{u}^2)^{-1}+\Omega^2{\H}\dot{u}^2\right)\B_1\,Z^{(i)} \,, \label{InvGeoDevBHi}
\end{align}
where $\dot{u}$ of the fiducial observer with given energy is given by (\ref{geod_eq_dot_u}) combined with (\ref{geod_eq_dot_r}), and $\Omega$ and $\H$ are given by (\ref{Omega_sol_gauge}) and (\ref{H_sol_gauge}), respectively. Differences from Newton-like deformations on an isotropic background directly reveal the quadratic gravity contributions encoded in the Bach tensor components $\B_1$ and $\B_2$. These discrepancies arise from the second lines in (\ref{InvGeoDevBH1}) and (\ref{InvGeoDevBHi}), respectively. In contrast to the Schwarzschild--(anti-)de Sitter solution, the anisotropic behavior given by $\Psi_2$, $\B_1$, and $\B_2$ survives even in the asymptotic region with $\Omega\rightarrow\infty$. Close to infinity they become independent of observer energy $E$. In particular, using the fully explicit form of (\ref{InvGeoDevBH1}) and (\ref{InvGeoDevBHi}) and making such a limit, we get
\begin{align}
\ddot{Z}^{(1)} =&
\frac{\Lambda}{3}\,Z^{(1)} + \frac{2\Lambda}{9}\,Z^{(1)} +\frac{4\Lambda}{9}\,Z^{(1)}=\Lambda Z^{(1)} \,, \label{InvGeoDevBH1_as}\\
\ddot{Z}^{(i)} =& \frac{\Lambda}{3}\,Z^{(i)} -\frac{\Lambda}{9}\,Z^{(i)}-\frac{\Lambda}{9}\,Z^{(i)}= \frac{\Lambda}{9}\,Z^{(i)} \,, \label{InvGeoDevBHi_as}
\end{align}
where the middle parts indicate particular contributions of the scalar curvature, Weyl tensor, and Ricci tensor, respectively.

\subsection{Alternative coordinates\label{SubSec:AltCoord}}

\begin{figure}[h]
    \centering
    \includegraphics[width=0.45\textwidth]{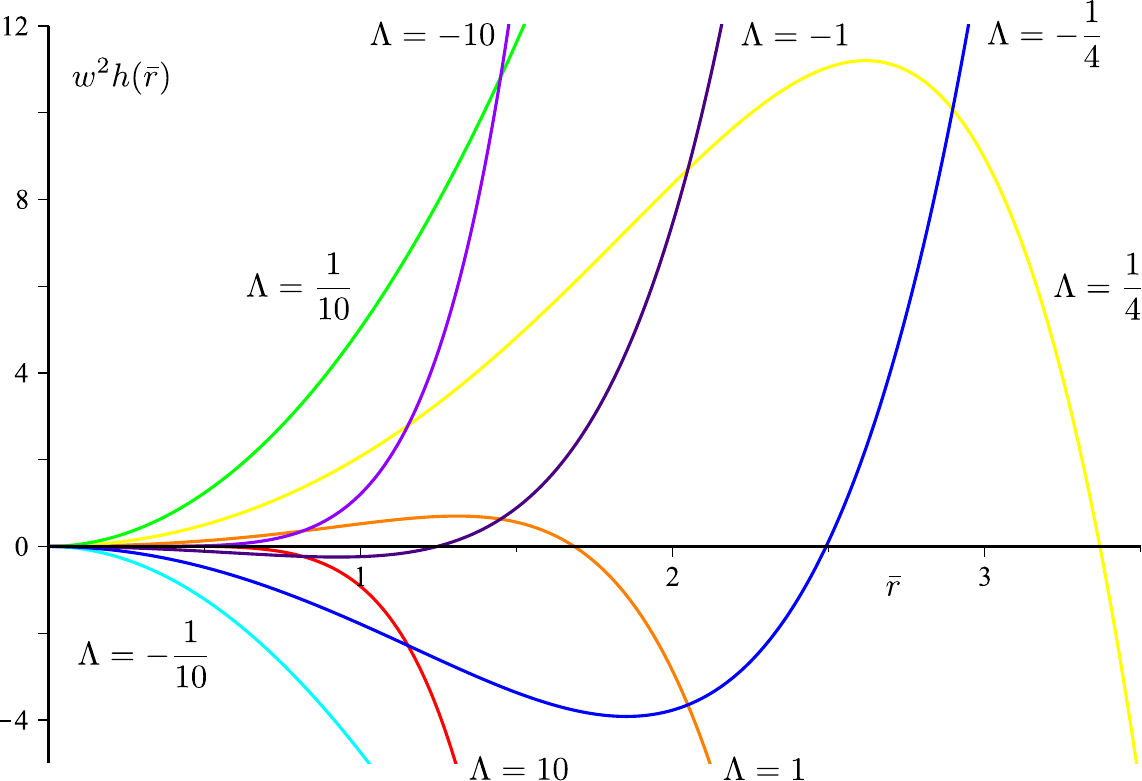}
    \caption{Visualisation of the metric function $h(\bar{r})$ scaled by the factor $w^2$ corresponding to the time-scaling freedom of the standard spherical coordinates (\ref{metric:SchwCoord}). }\label{fig:h_Schw}
\end{figure}

\begin{figure}[h]
    \centering
    \includegraphics[width=0.45\textwidth]{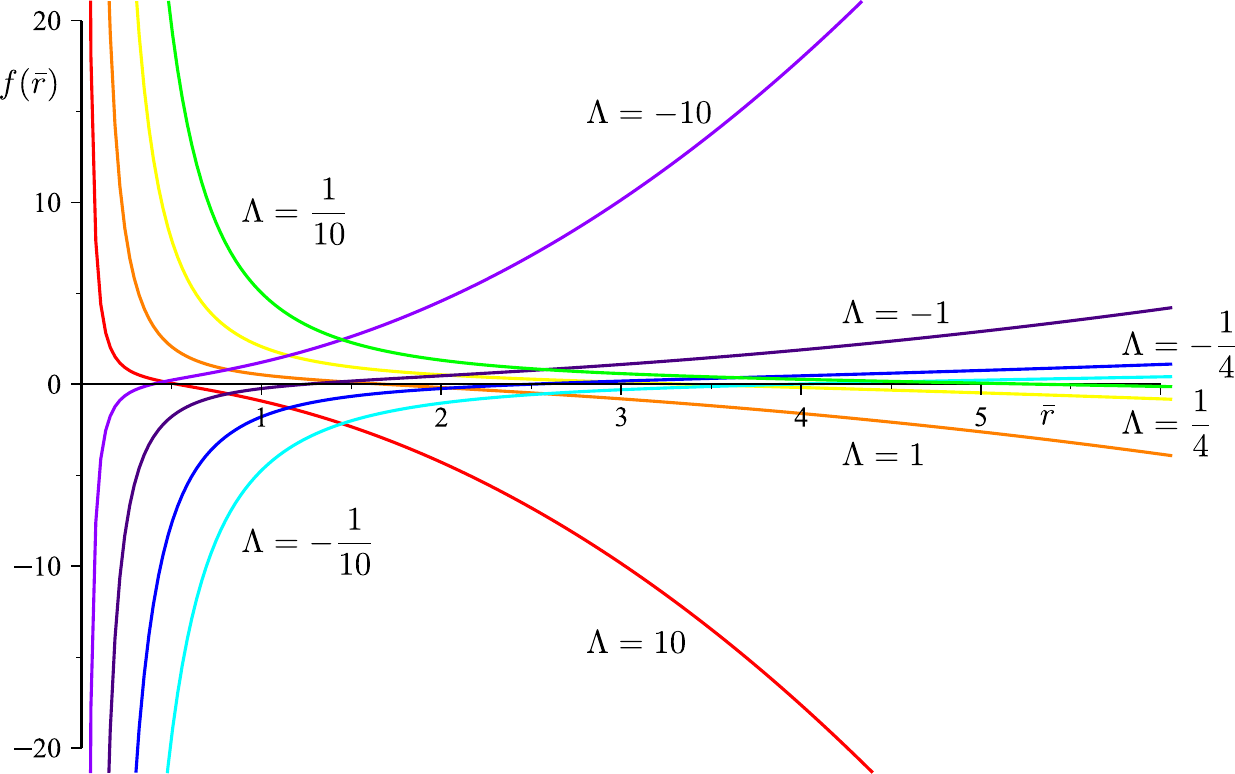}
    \caption{Visualisation of the metric function $f(\bar{r})$ of the standard spherical line element (\ref{metric:SchwCoord}). }\label{fig:f_Schw}
\end{figure}

For the standard spherically symmetric parameterization (\ref{metric:SchwCoord}) the relation (\ref{trans:SchwToConfKundt}) implies $\bar{r}=\Omega(r)$, which preserves a clear geometric interpretation of the coordinate~$\bar{r}$ measuring the surface radius and determining the position of the horizon as $\bar{r}_h=\Omega_h$. Using (\ref{rcehf}) the metric coefficients $g_{tt}$ and $g_{rr}$ become
\begin{align}
h(\bar{r}) &= -\frac{\bar{r}^{2}}{w^2}\left(\frac{\Lambda}{9}\,\bar{r}^4-\frac{1}{7}\,\bar{r}^2-\frac{24}{49\Lambda}\right)\,, \\
f(\bar{r}) &= -\frac{1}{\bar{r}^{2}}\left(\frac{\Lambda}{9}\,\bar{r}^4-\frac{1}{7}\,\bar{r}^2-\frac{24}{49\Lambda}\right)\,,
\end{align}
where the gauge parameter $w$ represents the scaling freedom of the time coordinate $t$. The typical behavior of these metric functions is plotted in Figures~\ref{fig:h_Schw} and \ref{fig:f_Schw}.


Finally, using the choice (\ref{nu_gauge}) for simplicity, the first expression in (\ref{trans:ConfKundtToRT}), and in addition requiring
\begin{equation}
r=0 \quad \leftrightarrow \quad \bar{r}=0 \,, \quad \leftrightarrow \quad \tilde{r}=0 \,,
\end{equation}
we get the coordinate relation
\begin{equation}
\tilde{r}=\frac{1}{3}w^2r^3 \,.
\end{equation}
Then the Robinson--Trautman form (\ref{metric:RT}) of spacetime (\ref{metric:ConfKundt}) with (\ref{Omega_sol}) and (\ref{H_sol}) is determined by
\begin{align}
\Omega(\tilde{r}) & = \left(3w\,\tilde{r}\right)^{1/3} \,,\\
H(\tilde{r}) & = \Lambda\tilde{r}^2 -\frac{1}{7}\frac{\left(3w\, \tilde{r}\right)^{4/3}}{w^2} -\frac{24}{49\Lambda}\frac{\left(3w\, \tilde{r}\right)^{2/3}}{w^2} \,.
\end{align}
Here, the area of surfaces with $u=\mbox{const}$ and $\tilde{r}=\mbox{const}$ is still proportional to $\Omega(\tilde{r})^2$, however the meaning of the coordinate $\tilde{r}$ itself is changed.

\section{Relation to previous results\label{Sec:Connection}}

As we have already mentioned, spherically symmetric spacetimes with the cosmological parameter have been extensively studied in quadratic gravity using conformal-to-Kundt coordinates (\ref{metric:ConfKundt}) and power series expansions~\cite{PravdaPravdovaPodolskySvarc:2021}. In particular, the metric functions were expressed as
\begin{align}
\Omega(r) =& \Delta^n   \sum_{i=0}^\infty a_i \,\Delta^{i}\,, \qquad \H(r) = \Delta^p \,\sum_{i=0}^\infty c_i \,\Delta^{i}\,, \label{rozvojcal_Omega0_H0}
\end{align}
where $\Delta\equiv r-r_0$ with $r_0$ being a real constant specifying the expansion center, while the leading powers defining particular subclasses are determined by specific values of $n$ and $p$ since the terms ${a_0}$ and ${c_0}$ were assumed to be non-vanishing.

The most interesting class $[n,\,p] =[0,\,1]$ was presented in Section V.B. of \cite{PravdaPravdovaPodolskySvarc:2021}, where the manifestly horizon-admitting spherically symmetric solution to quadratic gravity was derived. Simultaneously, the so-called Bach parameter $b$ was introduced as
\begin{equation}
b=\frac{1}{3}\left(c_1-2+\Lambda a_0^2\right) \,. \label{Bach_param_def}
\end{equation}
Then a specific choice of the gauge (\ref{GaugeTrans}) was made such that for $b=0$ the solution explicitly becomes the Schwarzschild--(A)dS spacetime (\ref{sol:SchwAdS_ConfKundtCoord}). Using such a gauge, the metric functions with $b\neq0$ expressed as expansions around $r_0=r_h$ are
\begin{widetext}
\begin{align}
\Omega(r) & = -\frac{1}{r}-\frac{b}{r_h}\sum_{i=1}^\infty\alpha_i\Big(\,\frac{r_h-r}{\X\,r_h}\Big)^i \,, \label{Omega_[0,1]}\\
\H (r) & = (r-r_h)\bigg[\,\frac{r^2}{r_h}-\frac{\Lambda}{3r_h^3}\left(r^2+rr_h+r_h^2\right)
+3b\,\X\,r_h\sum_{i=1}^\infty\gamma_i\Big(\,\frac{r-r_h}{\X\,r_h}\Big)^i\,\bigg] \,, \label{H_[0,1]}
\end{align}
where
\begin{align}
\X &\equiv  1-\frac{\Lambda}{r_h^2}\,, \quad \alpha_1 \equiv 1\,,\quad \gamma_1=1\,, \quad \gamma_2 = \frac{1}{3}\Big[4-\frac{1}{r_h^2}\Big(2\Lambda+\frac{1}{2k}\Big)+3b\Big] \,,
\end{align}
and the remaining coefficients $\alpha_l, \gamma_{l+1}$ for ${l \ge 2}$ are given by the recurrent relations (with ${\alpha_0=0}$)
\begin{align}
&\alpha_{l}= \, \frac{1}{l^2}\Big[-\frac{2\Lambda}{3r_h^2}\,\sum_{j=0}^{l-1}\sum_{i=0}^{j}\big[\alpha_{l-1-j}\X^j+\big(\X^{l-1-j}
+b\,\alpha_{l-1-j}\big)\big(\alpha_i \X^{j-i}+\alpha_{j-i}(\X^i+b\,\alpha_{i})\big)\big]-\frac{1}{3}\alpha_{l-2}(2+\X)\X(l-1)^2 \nonumber\\
& \hspace{14.0mm} +\alpha_{l-1}\big[\frac{1}{3}+(1+\X)\big(l(l-1)+\frac{1}{3}\big)\big]
-3\sum_{i=1}^{l}(-1)^i\,\gamma_i\,(\X^{l-i}+b\,\alpha_{l-i})\big(l(l-i)+\frac{1}{6}i(i+1)\big)\Big]\,, \nonumber\\
&\gamma_{l+1}= \, \frac{(-1)^{l}}{kr_h^2\,(l+2)(l+1)l(l-1)}\sum_{i=0}^{l-1}\big[\alpha_i \X^{l-i}
+\alpha_{l-i}\big(\X^i+b\,\alpha_i\big) \big](l-i)(l-1-3i) \, \quad \hbox{for}\quad l\geq2\,.
\label{alphasIIbgeneral}
\end{align}
\end{widetext}
This spacetime is fully characterized by the theory coupling constants, the cosmological parameter $\Lambda$, the horizon position $r_h$, and the Bach parameter $b$. These parameters are fully independent and unconstrained. This therefore means that the solution (\ref{Omega_sol_gauge}) and (\ref{H_sol_gauge}) derived in this paper can also be rewritten in the form (\ref{Omega_[0,1]}) and (\ref{H_[0,1]}), however, with all parameters strictly determined.

Here, we identify the solution (\ref{Omega_sol_gauge}) and (\ref{H_sol_gauge}) in such a form. More precisely, we show that it can also be rewritten within alternative $[n,\,p]$ classes, differing by a specific choice of a point around which the expansion is constructed. This could further shed light on the connections between various solution classes on a general level of Frobenius-like expansions in \cite{PravdaPravdovaPodolskySvarc:2021} and the convergence of particular power series.

First, the metric functions (\ref{Omega_sol_gauge}) and (\ref{H_sol_gauge}) are directly in the form of power series (\ref{rozvojcal_Omega0_H0}) within the class $[n,\,p] =[0,\,0]$. Denoting $\Omega_0 \equiv \Omega(r_0)=wr_0+v$, this expansion is determined by
\begin{align}
a_0 = \Omega_0 \,, \qquad a_1=w \,,
\end{align}
and
\begin{align}
c_0 =& \frac{1}{w^2}\left(\frac{\Lambda}{9}\,\Omega_0^4-\frac{1}{7}\,\Omega_0^2-\frac{24}{49\Lambda}\right) , \label{c_0_class_00} \\
c_1 = & \frac{2\Omega_0}{w}\left(\frac{2\Lambda}{9}\,\Omega_0^2-\frac{1}{7}\right) \,, \label{c_1_class_00}\\
c_2 =& \frac{2\Lambda}{3}\,\Omega_0^2-\frac{1}{7} \,, \\
c_3 =& \frac{4\Lambda}{9}w\,\Omega_0\,, \\
c_4 =& \frac{\Lambda}{9}w^2 \,, \label{c_4_class_00}
\end{align}
while all the remaining coefficients vanish. 

Simultaneously, we may express our solution as the near-horizon expansion, i.e., $r_0\rightarrow r_h$ is the root of~(\ref{H_sol_gauge}) and thus satisfies $\H(r_h)=0$, while $\Omega_0\rightarrow\Omega_h$. For $r_h$, we have explicitly obtained (\ref{cK_r_hozinot_explicit}) and $\Omega_h$ is given by (\ref{horizon_cK_coord_Omega}). Since the coefficient (\ref{c_0_class_00}) vanishes in such a case, to be consistent with the notation of \cite{PravdaPravdovaPodolskySvarc:2021} where $c_0$ is assumed to be non-vanishing, we have to re-label the coefficients (\ref{c_1_class_00})--(\ref{c_4_class_00}) as $c_i\rightarrow c_{i-1}$. The leading powers then become $[n,\,p] =[0,\,1]$ and the metric functions (\ref{Omega_sol_gauge}) and (\ref{H_sol_gauge}) can be rewritten in the form of (\ref{rozvojcal_Omega0_H0}) as
\begin{align}
    \Omega = &\Omega_h + w\Delta \,,  \label{Omega_near_horizon_cK} \\
    \mathcal{H} =& \Delta\left[
    \frac{2\Omega_h}{w}\left(\frac{2\Lambda}{9}\,\Omega_h^2-\frac{1}{7}\right)
    +\left(\frac{2\Lambda}{3}\,\Omega_h^2-\frac{1}{7}\right)\Delta \right. \nonumber\\
   & \left. \qquad +\frac{4\Lambda}{9}w\,\Omega_h\,\Delta^2 +\frac{\Lambda}{9}w^2\,\Delta^3\right], \label{H_near_horizon_cK}
\end{align}
where $\Delta = r - r_h$. The coefficients that multiply the specific powers of $\Delta$ represent $a_0$, $a_1$ and $c_0$, $c_1$, $c_2$, $c_3$ of the expansion in class $[n,\,p] =[0,\,1]$, respectively. Now, let us rewrite this solution in a similar form to (\ref{Omega_[0,1]}) and (\ref{H_[0,1]}). First, we have to identify the Bach parameter $b$, which is given by (\ref{Bach_param_def}) with the above coefficients $a_0$ from (\ref{Omega_near_horizon_cK}) and $c_1$ from (\ref{H_near_horizon_cK}) with (\ref{horizon_cK_coord_Omega}). We get
\begin{equation}
    b =5\left(-\frac{1}{7}+\frac{\Lambda}{9}\Omega_h^2\right) =\frac{5}{42}\left(-3+\sgnL \sqrt{105}\right) \,.
\end{equation}
The explicit value $r_h$ of the coordinate $r$ that determines the horizon is given by (\ref{cK_r_hozinot_explicit}).

Subsequently, we can artificially reconstruct the Schwarzschild--(A)dS background (\ref{sol:SchwAdS_ConfKundtCoord}) in (\ref{Omega_near_horizon_cK}) using 
\begin{align}
\frac{1}{r}= \frac{1}{r_h}\sum_{i=0}^{\infty}\left(\frac{r_h-r}{r_h}\right)^i \,,
\end{align}
where the radius of convergence of this series restricts the range of $r$ where such a solution is valid. This leads to
\begin{align}
\Omega(r) & = -\frac{1}{r}+\frac{1}{r_h}\sum_{i=0}^\infty\tilde{\alpha}_i\Big(\,\frac{r_h-r}{\,r_h}\Big)^i \,, \label{Omega_[0,1]_reconstructed}
\end{align}
where
\begin{equation}
\tilde{\alpha}_0=1+r_h\Omega_h \,, \ \ \tilde{\alpha}_1=1-r_h^2w \,, \ \ \tilde{\alpha}_i=1 \ \mbox{for} \ i\geq2 \,.
\end{equation}
An analogous reshuffling of the terms can also be performed in (\ref{H_near_horizon_cK}) to obtain
\begin{align}
 \mathcal{H}(r) = (r-r_h)\left[\frac{r^2}{r_h}\right. & -\frac{\Lambda}{3r_h^3}\left(r^2+rr_h+r_h^2\right) \nonumber \\
 & \left. +\sum_{i=0}^\infty\tilde{\gamma}_i(r-r_h)^i \right], \label{H_[0,1]_reconstructed}
\end{align}
where $\tilde{\gamma}_i=0$ for $i\geq4$ and
\begin{align}
\tilde{\gamma}_0= &\frac{\Lambda}{r_h}-r_h+\frac{2\Omega_h}{w}\left(\frac{2\Lambda}{9}\,\Omega_h^2-\frac{1}{7}\right) , \\
\tilde{\gamma}_1= &\frac{\Lambda}{r_h^2}-1+\left(\frac{2\Lambda}{3}\,\Omega_h^2-\frac{1}{7}\right),  \\
\tilde{\gamma}_2= &\frac{\Lambda}{3r_h^3}-\frac{1}{r_h}+\frac{4\Lambda}{9}w\,\Omega_h\,, \\
\tilde{\gamma}_3= &\frac{\Lambda}{9}w^2\,.
\end{align}

\begin{figure}[h]
    \centering
    \includegraphics[width=0.45\textwidth]{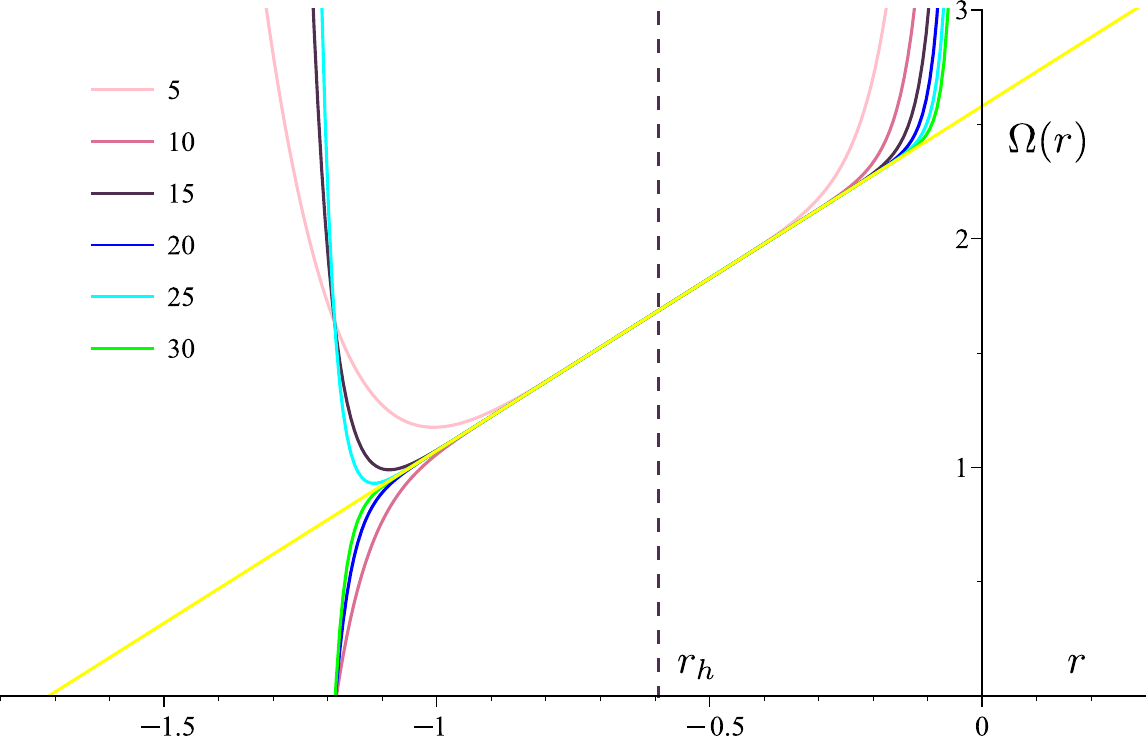} 
    \caption{Comparison of the exact metric function $\Omega(r)$, yellow straight line, given by (\ref{Omega_near_horizon_cK}) with the gauge (\ref{comp_gauge}) and its near-horizon expression in terms of the power series (\ref{Omega_[0,1]}), where the colours correspond to a specific order of the expansion. The dashed line indicates the position of the horizon $r_h$.}\label{fig:Omega_series_cK}
\end{figure}

The gauge freedom is still present in the above expressions (\ref{Omega_[0,1]_reconstructed}) and (\ref{H_[0,1]_reconstructed}) through $r_h$; however, $\Omega_h$ is gauge-independent. To compare the solution with (\ref{Omega_[0,1]}) and (\ref{H_[0,1]}) the gauge has to be fixed. In particular, conditions $\tilde{\alpha}_0=0$ and $\tilde{\alpha}_1=-b\rho^{-1}$ must be met to be consistent with (\ref{Omega_[0,1]}). The corresponding choice is
\begin{equation}
v = \Omega_h\,\frac{2\Lambda\Omega_h^2 - b - 2}{\Lambda\Omega_h^2 - 1}\,, \quad w = \Omega_h^2\,\frac{\Lambda\Omega_h^2 - b - 1}{\Lambda\Omega_h^2 - 1} \,. \label{comp_gauge}
\end{equation}
Then, all the remaining coefficients coincide with those given by the generic expressions (\ref{alphasIIbgeneral}). The power series expansion (\ref{Omega_[0,1]}), or equivalently (\ref{Omega_[0,1]_reconstructed}) with (\ref{comp_gauge}), of the metric function $\Omega$ compared to the exact solution (\ref{Omega_near_horizon_cK}) is visualized in Figure~\ref{fig:Omega_series_cK}.

To summarize, from this perspective, the conformal-to-Kundt spacetime (\ref{metric:ConfKundt}) with the quadratic gravity solution (\ref{Omega_sol_gauge}) and (\ref{H_sol_gauge}) described in Section~\ref{Sec:solution} clearly represents the Schwarzschild--Bach--(A)dS black-hole \cite{PravdaPravdovaPodolskySvarc:2021} with a very specific coupling of its parameters. Thus, all the previous results can be applied directly. Vice versa,  this solution may give a hint on relations between various classes of spherically symmetric spacetimes in quadratic gravity where the parameters connection is not straightforward. Simultaneously, the examples of the alternative forms of $\Omega(r)$ emphasize that one must be cautious about the choice of the gauge and the convergence of the solution in terms of the power series itself.

\section{Conclusion}

In this work, we have identified and analyzed a fully explicit static spherically symmetric solution in quadratic gravity (\ref{eq:quadratic_gravity_action}) under the assumption of constant scalar curvature (\ref{eq:trace_general}). Using the conformal-to-Kundt line element~(\ref{metric:ConfKundt}), the metric functions that solve the field equations (\ref{eq:field_eq}) become (\ref{Omega_sol_gauge}) and (\ref{H_sol_gauge}) together with a special coupling of the quadratic–gravity parameters (\ref{cont_relation}), see Section~\ref{Sec:solution}. From a geometric perspective, the solution contains a central curvature singularity accompanied by either a black‑hole horizon or a cosmological horizon, depending on the sign of the cosmological parameter $\Lambda$. Seen by the timelike observer, the spacetime is thus split into two regions. These aspects are illustrated by an explicit evaluation of the curvature invariants and related analysis of the behavior of geodesics. 

Although this spacetime belongs to the previously studied classes of spherical spacetimes in quadratic gravity, typically described using (approximative) methods such as infinite power series, continued fractions, or numerical simulations, its explicit closed form had not been recognized before. Here, we demonstrate its relation to the extensively studied Frobenius expansions and show the treacherousness hidden in the gauge choice and the Schwarzschild--(anti-)de Sitter background identification possibly restricting the convergence of the series.

Despite the highly constrained and physically limited nature of the described model, it reveals a complete global geometric structure and admits a very simple mathematical form that is typically obscured in more general, however, typically approximate and enormously complicated, solutions to quadratic gravity. As such, the solution may provide a test bed for studies of more realistic situations through a wide range of methods.

\begin{acknowledgments}

{\v S.} K. was supported by Charles University Grant Agency project No. GAUK 425425 and by the Charles University Research Center Grant No. UNCE24/SCI/016. {\v S.} K. and D. K. acknowledge support by the INTER-COST LUC25003 project of the INTER-EXCELLENCE II program of the Czech Ministry of Education, Youth and Sports and the contribution of the COST Action CA23130. R. {\v S}. is grateful for the support of the Czech Science Foundation Grant No. GACR 23-05914S.

\end{acknowledgments}

\end{document}